\documentclass[prd,12pt,onecolumn,nofootinbib,amsmath,amssymb,aps,floatfix]{revtex4-1}
\usepackage{amssymb,amsmath,graphicx,color}
\usepackage[linktoc=page]{hyperref}
\usepackage{subfigure}
\usepackage{amsfonts}
\newcommand{\beq}{\begin{equation}}
\newcommand{\eeq}[1]{\label{#1}\end{equation}}
\newcommand{\bea}{\begin{eqnarray}}
\newcommand{\eea}[1]{\label{#1}\end{eqnarray}}

\begin{document}
\title{Observable Supertranslations}
\author{Raphael Bousso}%
 \email{bousso@lbl.gov}
\affiliation{Center for Theoretical Physics and Department of Physics\\
University of California, Berkeley, CA 94720, USA 
}%
\affiliation{Lawrence Berkeley National Laboratory, Berkeley, CA 94720, USA}
\author{Massimo Porrati}%
 \email{mp9@nyu.edu}
\affiliation{Department of Physics and CCPP\\ New York University, New York NY 10003, USA}
\begin{abstract}
We show that large gauge transformations in asymptotically flat spacetime can be implemented by
sandwiching a shell containing the ingoing hard particles between two finite-width shells of soft gauge excitations.
Integration of the graviton Dirac bracket implies that our observable soft degrees of freedom obey the algebra imposed by Strominger et al. on unobservable boundary degrees of freedom. Thus, we provide both a derivation and an observable realization of this algebra. 

We recently showed that soft charges fail to constrain the hard scattering problem, and so cannot be relevant to the black hole information paradox. By expressing the BMS algebra in terms of observable quantities, the present work shows that this conclusion was not an artifact of working with strictly zero frequency soft modes. The conservation laws associated with asymptotic symmetries are seen to arise physically from free propagation of infrared modes. 
\end{abstract}
\maketitle
\tableofcontents

\section{Introduction}

For gauge theories with massless sources in asymptotically flat spacetime, the description of scattering requires an asymptotic gauge choice. In Maxwell theory, this is a gauge potential near infinity. This choice can affect the relative change of phase, for example, as two asymptotic components of a particle's wavefunction propagate from different angles to the scattering region. Thus it can affect the particle's amplitude in the scattering region and the outcome of the scattering experiment. Here we will largely focus on the case of gravity, where an asymptotic coordinate choice must be made.

In $D>4$ spacetime dimensions, it is possible to fix the asymptotic gauge potential permanently to a fiducial value, by working at sufficiently large radius. However, in $D=4$, the gauge is continually altered, at leading order in $1/r$, by the very particles that appear in the ingoing or outgoing radiation. This is the origin of the asymptotic symmetry groups corresponding to large gauge transformations, which transform the physical scattering data to a new asymptotic gauge. 

In massless Maxwell theory, the asymptotic gauge group is a $U(1)$ at every angle. In gravity, it is the Bondi-van der Burg-Metzner-Sachs (BMS) group~\cite{BMS,Sachs}. BMS transformations include supertranslations, which can be interpreted as angle-dependent time translations accompanied by $O(1/r)$ deformations of asymptotic spheres.

Strominger et al.~\cite{Str13,HeLys14} have proposed a symplectic structure such that asymptotic gauge transformations are generated by an infinite set of conserved charges, $Q=Q_H+N$. The first term is the total ``hard'' charged flux at every angle. The second term is an infinite ``memory'': the integral, over all of time, of the electromagnetic radiation or the Bondi news. 

The fact that these two terms are only conserved together means that the hard and soft sectors mix dynamically during a scattering process. This would seem to imply that information is lost when only the hard data are considered. Thus, it was suggested that the asymptotic gauge groups may be relevant to the black hole information paradox~\cite{HPS,HPS2}. 

We recently showed that this is not the case~\cite{BouPor17}. (See Ref.~\cite{MirPor16} for closely related earlier work, and Refs.~\cite{BouHal16,gabai,averin,gomez,panchenko,shiu,giddings17,semenoff17} for other work on this issue.) 
Our argument in~\cite{BouPor17} is particularly simple and direct, because it obtains already at the classical level. We found that the mixing of soft and hard degrees of freedom is entirely due to contributions of the long-range field of the hard particles to the soft charge. 

The hard-to-soft contributions are constrained by gauge invariance and can be isolated by a canonical transformation. Crucially, they contain no information independent of the hard data, except for an undetermined infrared homogeneous solution of the wave equations. Indeed, sufficiently long-wavelength excitations can always be added to a hard scattering problem without affecting its dynamics. (This is why we can operate particle accelerators without monitoring the cosmic microwave background.) The infinite number of new conservation laws identified in Refs.~\cite{Str13,HeLys14,HPS,HPS2} properly reflect the free propagation of this soft radiation through the spacetime~\cite{BouPor17}.

\subsection*{Summary and Outline} 
In the present paper, we focus on a different problem: we argue that using standard definitions~\cite{Ash81,Str13}, the soft variables are not observable even in principle. We introduce alternative variables which are observable, and we show that the algebra and conservation laws imposed by Strominger~\cite{Str13} can then be derived from standard Poisson brackets and physical considerations. In other words we show that the mathematical charges defined by Strominger et al. provide a good approximation to physical quantities. The fact that such variables exist is highly nontrivial, and without them, a discussion of soft modes would be operationally meaningless from the start. (However, our earlier conclusion is unaffected: the hard and soft sectors decouple, and the asymptotic gauge group has no bearing on the black hole information problem.) We focus on the gravitational case without matter; the Maxwell case is completely analogous, and the addition of matter is trivial.
In Sec.~\ref{sec-bob} we introduce an asymptotic observer, Bob, who operates a network of freely floating gravitational wave ``guns'' and detectors (i.e., test masses) occupying a sphere of very large radius $r$. We illustrate the notion of Bondi frames and BMS supertranslations in terms of the real-world problems faced by Bob when he attempts to carry out a scattering experiment. Bob's network is deformed by the ingoing and outgoing radiation. The test masses can be fixed to a fiducial position (e.g., a round sphere) at one time, but they will not maintain this shape. The hard scattering data for a given physical problem depends on this gauge choice. Specifically, it affects the proper times when bursts of radiation must be injected from various angles, and when they are expected to come out. Therefore, if a theorist wishes to hand Bob a well-posed scattering problem, a gauge choice must be specified along with the prescribed in-state and the predicted out-state.

In Sec.~\ref{sec-q} we review standard definitions: the soft gravitational memory $N$ which contributes to the generator $Q=Q_H+N$ of BMS supertranslations; its conjugate variable, the deformation $C$; and the commutation relations that are imposed upon $N$ and $C$. Next, we argue that both $C$ and $N$ are unobservable. For $C$ this is obvious, since it corresponds to a metric component and so depends on an arbitrary coordinate choice~\cite{BouHal16}. 

For the gravitational memory $N$, the argument is more subtle. We use only a weak necessary condition for a quantity to be observable in asymptotically flat space: that it must be possible to measure (or produce) it to any specified accuracy $\epsilon$ by a physical operation lasting some finite time (that may diverge as $\epsilon\to 0$). We show that $N$---and hence, $Q$---fails to satisfy this criterion, if $N$ is defined as an infinite-time memory or strict zero mode. At any finite time, the measured gravitational memory can differ by an arbitrarily large amount from the infinite-time integral that defines $N$. This is true even if the total energy of the state is known and has been measured to optimal precision.

Finally, we consider the conservation laws imposed on $Q$ and on $C$ in~\cite{Str13,HeLys14}. We review our earlier result~\cite{BouPor17} that they imply factorization of the soft and hard sectors after a simple canonical transformation. We note that if the soft sector did not decouple, information would be effectively lost (even in the Maxwell case, i.e., without black holes; and even classically), due to the unobservability of the variables $N$ and $C$. 

In Sec.~\ref{sec-cno}, we define $C$ and $N$ as the (observable) memories of certain large but finite time intervals. The bracket algebra of~\cite{Str13,HeLys14} need not be imposed, but can instead be derived from the Dirac bracket of the Bondi news. Physically, $C$ dials a Bondi frame for Bob's test masses, and $Q=Q_H+N$ generates changes of this frame (BMS supertranslations). The conservation of $C$ and $Q$ is also not imposed. It arises as a consequence of the free propagation of long-wavelength excitations, together with gauge invariance.

\section{Bondi Frames and the Scattering Problem}
\label{sec-bob}

The scattering of massless charged particles in four spacetime dimensions requires specification both of their flux and of an asymptotic gauge choice. This is most easily explained for the case of gravity, where the charged particles and the gauge bosons both correspond to gravitons. It can be discussed at a classical level. 

\subsection{A Gravitational Wave Scattering Experiment}

Alice has solved a scattering problem. As a theorist, she likes to work directly on the conformal boundaries of spacetime. Because she considers only massless particles, the relevant portions are past and future null infinity, ${\cal I}^-$ and ${\cal I}^+$. Each portion is topologically $\mathbf{R}\times \mathbf{S}^2$. Alice has specified a flux of gravitational waves entering the spacetime from ${\cal I}^-$, as a function of advanced time and solid angle, $N^-_{AB}(v,\bar\vartheta)$. She has computed the resulting outgoing flux on ${\cal I}^+$ in terms of retarded time and angle, $N_{AB}(u,\vartheta)$. 

\begin{figure}[t]
\includegraphics[width=0.7 \textwidth]{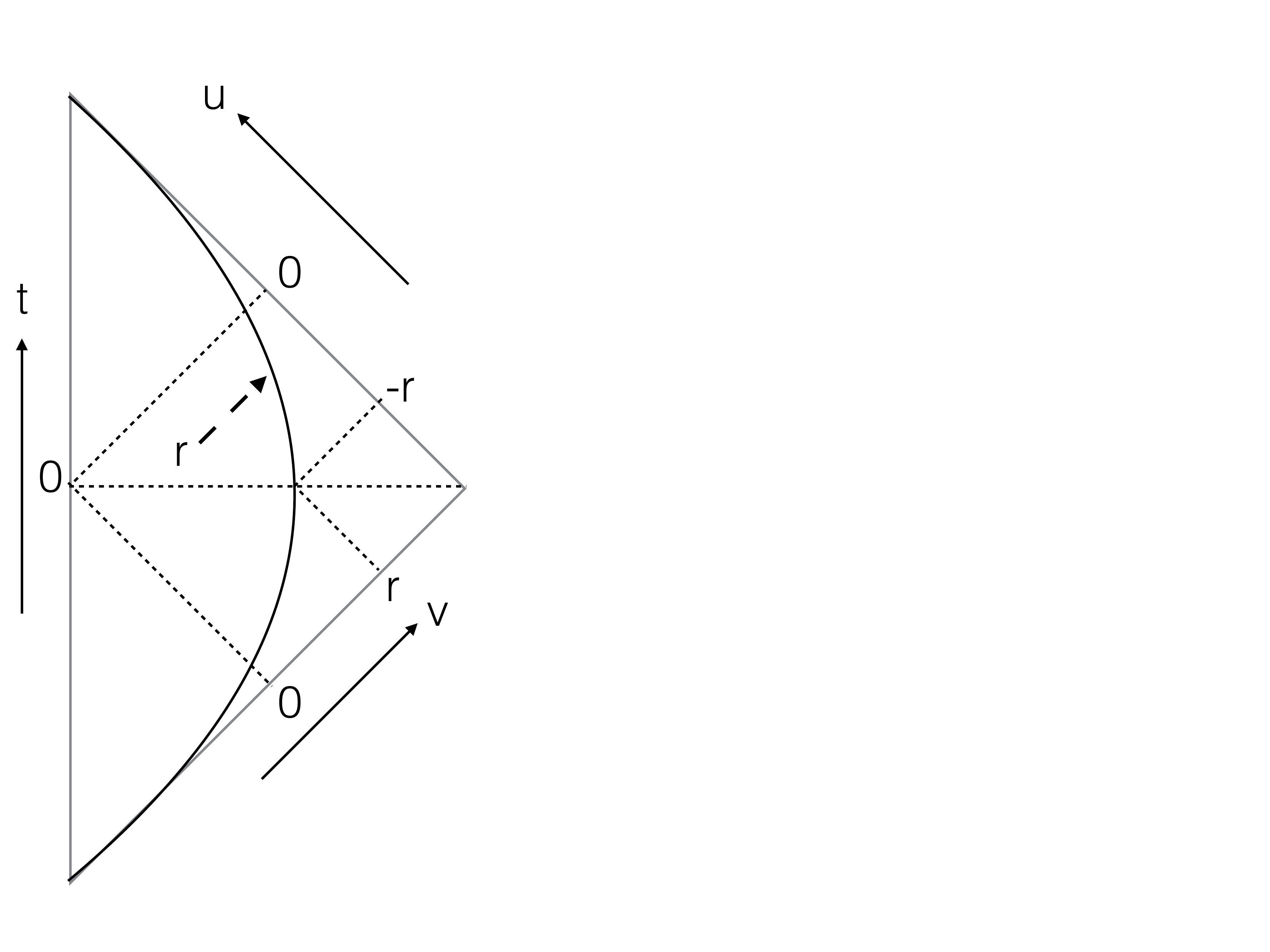}
\caption{Penrose diagram of a gravitational scattering process in asymptotically flat spacetime. The hard ingoing and outgoing particles are localized near $v=0$ and $u=0$ (left dashed wedge). Alice formally predicts an out-state on $\mathcal{I}^+$ from an in-state on $\mathcal{I}^-$ (top and bottom edges). Bob has only finite time, so he occupies a sphere of finite radius $r$ with freely falling equipment that produces the in-state and records the out-state. In $D=4$, Bob's sphere suffers deformations during the process.}
\label{fig-abc}
\end{figure}

Experimentalist Bob would like to perform the scattering experiment Alice has computed. Bob lives in the physical spacetime, not on the conformal boundary. So he populates a large two-sphere with freely falling gravitational wave emitters and detectors carrying synchronized clocks showing time $t$. He chooses the sphere's radius $r$ much greater than the duration of the ingoing and outgoing flux, so that there is a clean time separation between the two eras; see
Figure~(\ref{fig-abc}). By a trivial global coordinate shift he can choose the ingoing and outgoing bursts to be approximately symmetric about $t=0$. Thus the ingoing radiation enters around $t\approx -r$ ($v\approx 0, u\approx -2r$) and the outgoing radiation reaches Bob at $t\approx r$ ($v\approx 2r$, $u=0$). 

Also, $r$ is chosen so large that the radial position of Bob's equipment does not change appreciably, despite the gravitational attraction towards the center during the experiment. (It is straightforward to check that this can be arranged.) It will take Bob a time at least of order $r$ to carry out a scattering experiment, and thus it will take a long time if we take $r$ large. We do not restrict the time duration, or $r$, except for the obvious criterion that it must be finite for the experiment to be possible. 

Thus, Bob should interpret Alice's input data $N^-_{AB}(v,\bar\vartheta)$ at ${\cal I}^-$ as an instruction to send in waves $N^-_{AB}$ orthogonally through his big round sphere at angle $\bar\vartheta$ at the clock time $t=v-r$. Later, Bob's detectors receive gravitational waves $N_{AB}$ coming out orthogonally through the sphere at angles $\vartheta$ and clock times $t=u+r$, which he can compare to Alice's prediction. 

In the presence of gravity, the spacetime will be deformed. During the experiment, therefore, his clocks cannot remain exactly synchronized, and the proper distance between neighboring detectors may change. 

In $D>4$, Bob's equipment can be made sensitive to the gravitational radiation only, and all other effects of gravity can be made irrelevant, by choosing $r$ large enough. In $D=4$, however, the fluxes $N^-_{AB}, N_{AB}$ distort the distant spheres occupied by Bob's detectors, at $O(1/r)$. We will now quantify this effect. 

\subsection{Metric Adapted to Inertial Detectors}
\label{sec-metric}

It can be shown that Bob's detectors follow geodesics with proper time $t=u+r$ and fixed $r,\vartheta$ in the following metric\footnote{Capital indices correspond to angular directions. We omit terms subleading in $r$.}
\begin{eqnarray}
 ds^2 & = & -du^2 - 2 du\, dr + r^2 \left(h_{AB}
+\frac{C_{AB}}{r}\right) d\vartheta^A d\vartheta^B\nonumber\\
 & & + D^AC_{AB}\, du\, d\vartheta^B \label{eq-bondi}~,
\end{eqnarray}
where $h_{AB}$ is the metric on the unit round two-sphere. Crucially, the deformation of the sphere depends on the outgoing radiation:
\begin{equation}
C_{AB}(u,\vartheta) = C_{AB}(u_0,\vartheta) + \int_{u_0}^u du\, N_{AB}(u,\vartheta)~,
\label{eq-newsdef}
\end{equation}
where the integration constant, and hence $C_{AB}$, is symmetric and traceless, $h^{AB} C_{AB}=0$ and $C_{AB}=C_{BA}$.

The above metric covers the entire asymptotic region, but the limit as $r\to\infty$ at fixed $u$ reaches only ${\cal I}^+$, not ${\cal I}^-$. To make contact with Alice's specification of an in-state on ${\cal I}^-$, we will use the above metric only for $t>0$, or $u>-r$, where $r$ is the size of Bob's sphere. This was anticipated by including only the outgoing flux in Eq.~(\ref{eq-newsdef}). 

Otherwise, for $t<0$, we shall use advanced Bondi coordinates. Bob's test masses follow geodesics with proper time $t=v-r$ and fixed $r,\bar\vartheta$ in the metric 
\begin{eqnarray}
 ds^2 & = & -dv^2 + 2 dv\, dr + r^2 \left(h_{AB}
+\frac{C^-_{AB}}{r}\right) d\bar\vartheta^A d\bar\vartheta^B\nonumber\\
 & & - D^AC^-_{AB}\, dv\, d\bar\vartheta^B \label{eq-bondiv}~,
\end{eqnarray}
where
\begin{equation}
C^-_{AB}(v,\bar\vartheta) = C^-_{AB}(v_0,\bar\vartheta) + \int_{v_0}^v dv\, 
N^-_{AB}(v,\bar\vartheta)~.
\label{eq-newsdefv}
\end{equation}
Obviously the two charts are related by 
\begin{equation}
v-r=t=u+r~.
\label{eq-uv}
\end{equation}
In addition we may choose to introduce a nontrivial relation between angles, so that $\vartheta$ and $\bar \vartheta$ actually denote antipodal angles:
\begin{equation}
\vartheta = \pi-\bar\vartheta~~,~~~\phi = \pi+\bar\phi~.
\label{eq-antip}
\end{equation}
This has the advantage that trivial scattering preserves the angle, $\vartheta \to \bar \vartheta$. We will drop the overbars below, with the understanding that this relation between advanced and retarded coordinates is always imposed. 

Next, we turn to the integration constants in Eqs.~(\ref{eq-newsdef}) and (\ref{eq-newsdefv}).

\subsection{Supertranslations}

Recall that each test mass at fixed $r$ carries physical labels $\vartheta$ and a clock showing proper time $t=u+r=v-r$. At constant time $t$, the masses occupy a sphere that is round up to a deformation of order $1/r$ relative to the unit round sphere.  By Eqs.~(\ref{eq-newsdef}) and (\ref{eq-newsdefv}), one cannot choose the deformations $C_{AB}$ ($C^-_{AB}$) to vanish at all times, if there is nonzero flux $N_{AB}$ ($N^-_{AB}$). However, the choice of the integration constants is free, and it corresponds to a freedom in Bob's choice of detector arrangement at some fiducial times $u_0$ and $v_0$.

For example, Bob can choose to populate a round sphere at the intermediate time $t=0$, after all the ingoing flux has gone in, and before any outgoing flux arrives. This corresponds to choosing the integration constants so that $C_{AB}(t=0)=0=C^-_{AB}(t=0)$ at all angles. But then by Eq.~(\ref{eq-newsdef}) the detectors will occupy a deformed sphere at early and late times.

This effect can be measured. At leading order, the proper distance between two detectors separated by a small angle $\delta\vartheta$ is $r \delta\vartheta$, where $\delta\vartheta\equiv (h_{AB} \delta\vartheta^A\delta\vartheta^B)^{1/2}$. The correction at order $r^0$ is time-dependent:
\begin{equation}
\delta L= r\delta\vartheta + \frac{1}{2\delta\vartheta} C_{AB}(u,\vartheta) d\vartheta^A d\vartheta^B~.
\label{eq-deltal}
\end{equation}
Thus, Bob can determine $C_{AB}(u,\vartheta)$ by monitoring the separation of test masses, and he can measure the flux using Eq.~(\ref{eq-newsdef}):
\begin{equation}
N_{AB}(u,\vartheta) = \partial_u C_{AB}(u,\vartheta)~.
\label{eq-nduc}
\end{equation}
In particular, Bob can measure outgoing gravitational memory, defined as any definite integral of $N_{AB}$ over some range of $u$. Analogously, he can measure ingoing gravitational memory at early times.

The {\em fact} that the asymptotic geometry is described to equal accuracy in $1/r$ by any choice of the integration constant in (\ref{eq-newsdef}) implies that there exists a diffeomorphism that relates all possible choices. It can be shown that $C_{AB}(u_0,\vartheta)$ can be written in terms of a single function on the sphere:\footnote{We take $C$ to have no $l=0$ or $l=1$ components since they are annihilated in any case by the derivatives. Naively, $C_{AB}(u_0,\vartheta)$ should correspond to two functions on the sphere, because $C_{AB}$ is symmetric and traceless. However, in fact the ``magnetic part'' of $C_{AB}$ vanishes in regions with $N_{AB}=0$. (This has been proven in linearized gravity~\cite{BieGar13} and partially shown at the nonlinear level. See~\cite{FlaNic15} for a detailed discussion.) We shall take $u_0$ (or $v_0$) to lie in such a region: either at early times (before the flux goes in), or at intermediate times (after it goes in but before it comes out), or at late times (after all the flux comes out).}
\begin{eqnarray} 
C_{AB}(u_0,\vartheta) & = & (-2D_A D_B + h_{AB}D_C D^C)\, C(\vartheta)~,
\label{eq-vacc}\\
C^-_{AB}(v_0,\vartheta) & = & (-2D_A D_B + h_{AB}D_C D^C)\, C^-(\vartheta)~.
\label{eq-vaccm}
\end{eqnarray} 

The vector field
\begin{equation}
\xi_f = f\partial_u - D_Af\,\frac{\partial_A}{r} + \frac{1}{2}D^2f \,\partial_r~.
\label{eq-xi}
\end{equation}
generates a diffeomorphism on the retarded coordinates, called a Bondi-van der Burg-Metzner-Sachs (BMS) supertranslation~\cite{BMS,Sachs}. In the new coordinates, the metric again takes the form of Eqs.~(\ref{eq-bondi}), (\ref{eq-newsdef}), and (\ref{eq-vacc}), but with
\begin{equation}
C(\vartheta) \to C(\vartheta)+f(\vartheta)~.
\label{eq-ccf}
\end{equation}
Similarly in advanced coordinates, the vector field
\begin{equation}
\xi^-_{f^-} = f^-\partial_v - D_Af^-\,\frac{\partial_A}{r} + \frac{1}{2}D^2f^-\, \partial_r~.
\label{eq-xim}
\end{equation}
generates a BMS supertranslation. In the new advanced coordinates, the metric again takes the form of Eqs.~(\ref{eq-bondiv}), (\ref{eq-newsdefv}), and (\ref{eq-vaccm}), but with
\begin{equation}
C^-(\vartheta) \to C^-(\vartheta)+f^-(\vartheta)~.
\label{eq-ccf}
\end{equation}

Recall that Bob's detectors are adapted to the metric (\ref{eq-bondi}) if they occupy a sphere with deformation $C(\vartheta)$ at the time $u_0$. Alice computed the out-state assuming a particular choice of $C$. Now suppose that Bob changes his mind about $C$, i.e., about his detector arrangement at $u_0$. How will this affect his description of the out-state?  

By Eq.~(\ref{eq-ccf}), any two choices of $C$ are connected by a BMS supertranslation $\xi_f$. Since $\xi_f$ is a diffeomorphism, Bob can use $\xi_f$ to determine how the flux function predicted by Alice, $N_{AB}(u,\vartheta)$, must transform. He simply regards the transformation of $C$ as part of the full diffeomorphism described by $\xi_f$, which must also be applied to $N_{AB}$. The flux function describes the flux at the conformal boundary, $r\to\infty$, so only the first term in Eq.~(\ref{eq-xi}) acts on $N_{AB}$.\footnote{Bob is working at large but finite radius. Strictly speaking, he should also implement the $\partial_A$ and $\partial_r$ terms in $\xi_f$. This corresponds to a finite transverse translation of the flux, and a modification of its physical amplitude, $N_{AB}/r$. Both are subleading and become physically irrelevant at large $r$, however.} Thus $\xi_f$ implements an angle-dependent time shift by $f$:
\begin{equation}
N_{AB}(u,\vartheta)\to N_{AB}(u+f,\vartheta)~.
\label{eq-nnf}
\end{equation}

To summarize, a final state at late times is described by $\{N_{AB}(u,\vartheta),C\}$, the flux function together with a choice of Bondi gauge $C$. Its physical properties remain unchanged under any diffeomorphism (\ref{eq-xi}), that is, under the simultaneous transformations (\ref{eq-ccf}), (\ref{eq-nnf}). Analogous statements hold for the ingoing flux and advanced coordinates. 

\subsection{Remarks}

We close this section with two remarks. First, we have not yet tied together the BMS transformations Bob can perform on the ingoing and outgoing radiation. Below we will consider introducing a symplectic structure that naturally links them, but this is optional. A priori, the two integration constants can be chosen independently. So long as Alice specifies an in-state in terms of $C^-(\vartheta)$ and $N^-_{AB}(v,\vartheta)$, and an out-state in terms of $C(\vartheta)$ and $N_{AB}(u,\vartheta)$, her prediction can be checked: it corresponds to a well-defined scattering experiment that Bob can perform.

Second, recall that we use the advanced metric (\ref{eq-bondiv}) to describe Bob's sphere at early times, $t<0$, the era of ingoing radiation. We use the retarded metric (\ref{eq-bondi}) to describe the outgoing radiation, $t>0$. Moreover, we required that Bob's guns and detectors follow geodesics at fixed $r, \vartheta$. This means that there could be a discontinuities at $t=0$. In principle, this presents no difficulty. It just means that Bob may have to change his clocks, or power his test masses with rocket engines that move them to a new position over some short time near $t=0$. Below we will often make choices that avoid this, for (Bob's) convenience.

\section{BMS Generators and Charges}
\label{sec-q}

We will now consider supertranslations as symplectic flow in a phase space of in-states, or of out-states. In this section, we will review standard definitions of soft variables that generate this flow. We will show that these standard quantities are not physically observable. 

We will also review identifications that can be imposed between variables at past and future infinity. These identifications have been interpreted as conservation laws~\cite{Str13,HeLys14}. They were recently shown to imply factorization of the soft sector~\cite{BouPor17}.

In Sec.~\ref{sec-cno}, we will provide an alternate definition of soft variables, which renders them observable. 

\subsection{Symplectic Structure and Supertranslation Charge}

We introduced the BMS supertranslation generator as a vector field $\xi_f$, in Eq.~(\ref{eq-xi}). By Eqs.~(\ref{eq-ccf}) and (\ref{eq-nnf}), $\xi_f$ acts on phase space as 
\begin{equation}
C\to C+f(\vartheta) ~,~~u\to u+f(\vartheta)~.
\end{equation}
$C(\vartheta)$ is the integration constant in the Bondi metric; see Eqs.~(\ref{eq-bondi}), (\ref{eq-newsdef}), and (\ref{eq-vacc}). The $u$-transformation is shorthand for angle-dependent time shifts of all dynamical fields, such as (\ref{eq-nnf}).  We now seek to obtain the same transformations as a symplectic flow in phase space.

The generator of angle-dependent translations, $f(\vartheta) \partial_u\,$, can be written as
\begin{equation}
Q_H[f] = \int_{{\cal I}^+} d^2 \vartheta\, f(\vartheta) \int_{-\infty}^\infty  du\, T_{uu} ~.
\end{equation}
where
\begin{equation}
T_{uu} = \frac{1}{32\pi G} N_{AB} N^{AB}+T^m_{uu},
\end{equation}
is the boundary stress tensor (energy flux of gravitational waves, per unit solid angle). In the following we will often omit the stress energy tensor of matter, $T^m_{uu}$, since none of our results depend on its presence.  The integral need not be taken strictly on the boundary but can be approximately evaluated at any sufficiently large finite radius. $Q_H$ will be called the {\em hard charge}. Analogous quantities can be defined on ${\cal I}^-$.

To obtain a generator of $C\to C+f$, one may formally introduce a conjugate variable $N$ that obeys the Dirac bracket
\begin{equation}
\{N(\vartheta),C(\vartheta')\} = 16\pi G\, \delta^2(\vartheta-\vartheta')~.
\label{eq-cncom}
\end{equation} 
$N$ and $C$ each commute with all hard variables, such as $N_{AB}$. Then the symplectic flow generated by the soft charge 
\begin{equation}
Q_S[f] = \frac{1}{16\pi G} \int d^2 \vartheta\, f(\vartheta)\, N(\vartheta)
\end{equation}
transforms $C\to C+f$ and leaves $N_{AB}(u,\vartheta)$ invariant. 

Thus the generator of supertranslations on phase space is
\begin{equation}
Q[f] = Q_H[f]+Q_S[f]~.
\end{equation}
This is the BMS supertranslation charge~\cite{Ash81,Str13,HeLys14}. 

Note that the value of $N$ on a given solution can be changed at will without affecting the value of $Q_H$, by acting with the phase space flow generated by $C$. In particular, this implies that the charge $Q$ can have nonzero value. Therefore, $Q$ connects different states in phase space~\cite{BanRey16}.

The hard charge $Q_H[f]$ is a total radiated energy, weighted over angles by $f$. It generates an obvious generalization of ordinary time translations. We have expressed $Q_H$ as an integral of the Bondi news at ${\cal I}^+$. But Bob can measure $N_{AB}$ at finite radius by Eq.~(\ref{eq-nduc}). Hence Bob can measure $T_{uu}$ and $Q_H$ at finite $r$.

We now turn to the question of how to interpret the new degree of freedom $N$ underlying the soft charge $Q_S$. We will show that $N$ can be realized in different ways, such that $C$ and $N$ are either both unobservable (in the next subsection), or both observable (in Sec.~\ref{sec-cno}).

\subsection{$C,N$ as Unobservable Degrees of Freedom}

We may regard $C$ as defining the metric components $C_{AB}$ at one particular cut, say at $u\to -\infty$. This is the viewpoint taken in Refs.~\cite{Str13,HeLys14}. Thus, $C$ measures the deformation of a coordinate sphere at constant $u,r$ at order $1/r$ relative to the round metric. Correspondingly $C^-$ is defined at the cut $v\to\infty$ in~\cite{Str13,HeLys14}.

Any metric component is obviously unobservable, since it can be changed by choosing different coordinate labels. It does not represent information intrinsic to the spacetime. A coordinate choice can be made physical by placing actual markers and clocks, as described for Bob's test masses above, and we will revisit this viewpoint below.
Here we regard $C$ as a metric component in (\ref{eq-bondi}). {\em As a metric component}, $C$ is not observable.

If $C$ is unobservable, then its canonically conjugate variable $N$ must be unobservable, too. For example, it would otherwise be possible to measure a wavefunction of $C$ by repeated measurements of $N$.

In Refs.~\cite{Str13,HeLys14}, the variable $N$ is identified with the zero mode of the Bondi news,
\begin{equation}
N \equiv -\int d^2 \vartheta \int_{-\infty}^\infty du\, D_A D_B N^{AB} ~, \label{eq-n}
\end{equation}
defined not as a limit, but strictly as zero frequency radiation. The Dirac bracket with $C$, Eq.~(\ref{eq-cncom}), is not derived but imposed. 

Naively it may seem that $N$, so defined, {\em is\/} observable. More precisely, one might expect a long finite observation of the Bondi news to provide an approximation to $N$, and that the approximation becomes increasingly accurate as the range of integration is increased. By the above remark this would imply that its conjugate $C$, too, is observable. 

Moreover, by Eq.~(\ref{eq-nduc}), any finite gravitational memory is manifestly independent of the choice of Bondi frame. Therefore the memory is not a property of the test masses but of the spacetime. Therefore, if $N$ defined in (\ref{eq-n}) were observable, then $C$ should be observable {\em by studying just the spacetime}. Yet, as noted above, such claims~\cite{HPS,HPS2} contradict the fact that coordinates are unobservable. (See also~\cite{Bou16,BouHal16,Bou16b}.)

The resolution of this apparent contradiction is that {\em no finite observation---and hence no observation---provides any approximation whatsoever} to $N$ as defined in (\ref{eq-n}). Suppose Bob has measured the memory $N_T$ accumulated over the finite, arbitrarily large interval $-T<u<0$. $N$ bears no relation to $N_T$ because radiation outside the interval can contribute an arbitrarily large memory of either sign. In fact, this can be done at arbitrarily low cost in energy, and so cannot be excluded by Bob even if he knows the total energy remaining in the spacetime at the time $u=0$ (which he cannot know to better than some finite precision). To see this, consider radiation in the unobserved interval $0<u<\tilde T$ with memory $N_{\tilde T}$. This can be realized by a single wavepacket of length $\tilde T$ with amplitude of order $N_{\tilde T}/\tilde T$. Its flux $T_{uu}$ scales as $(N_{\tilde T}/\tilde T)^2$, so the total unobserved energy scales as $N_{\tilde T}^2/\tilde T$. We can make this as small as we wish by taking $\tilde T$ large at fixed $N_{\tilde T}$. (Note that this argument does not apply to all observables, but only to those of arbitrarily low energy. For example, $Q_H$ can be measured in finite time with arbitrary precision. All Bob has to do is to wait until the total energy that he sent in has come out, to the corresponding precision.) 

In this respect, the charge $N$ is similar to the ``winding'' charge $Q$ of a 1+1 massless scalar $\phi$  in Minkowski space. The charge is obtained by integrating the current $J^\mu=\epsilon^{\mu\nu}\partial_\nu \phi$
\beq
Q=-\int_{-\infty}^{+\infty} dx \partial_x \phi =- \phi(+\infty) +\phi(-\infty).
\eeq{mp-scal1}
While vanishing for a free scalar, this charge can be nonzero if the scalar is coupled to external sources. An example is 
the  interaction $L_I=\partial_\mu\phi V^\mu$ coupling $\phi$ to an external vector $V^\mu$.  The charge of a finite
interval $x\in[-L,L]$, $Q[L]=-\int_{-L}^{L} dx \partial_x \phi $ is unrelated to $Q$, for the very same reasons given for $N$. For instance, notice that $Q[L]$ can be nonzero also in the absence of external sources. 

We conclude that with the definition (\ref{eq-n}), $N$ (and hence $Q_S$) is unobservable even in principle. This implies that the total charge $Q$ can be given any value at fixed $Q_H$ by changing the expectation value of $N$. In particular, this means that $Q$ can be nonzero. Hence, $Q$ generates formally nontrivial---though completely unobservable---transformations in phase space.

\subsection{Conservation Laws}

We now consider how the soft variables at ${\cal I}^+$ and ${\cal I}^-$ are related. We impose that the total charge $Q$ and the soft variable $C$ are both conserved:
\begin{equation} 
Q_H(\vartheta)+N(\vartheta) = Q_H^-(\vartheta)+N^-(\vartheta)~,~~~C(\vartheta) = C^-(\vartheta)~,
\end{equation}
where we recall that $\vartheta$ on ${\cal I}^+$ is antipodally related to $\vartheta$ on ${\cal I}^-$. 

It is important to stress that this antipodal conservation law is purely formal, since neither $C$ nor $N$ can be observed when defined as above. $Q_H$ of course can be observed, but $Q=Q_H+N$ cannot. (Any sum of quantities that includes at least one unobservable term is unobservable.) 

Using these conservation laws, we recently showed~\cite{BouPor17} that the soft sector decouples entirely from the hard scattering problem, after a simple canonical transformation

\bea
N(\vartheta) \rightarrow N^{D}(\vartheta) &=&  N(\vartheta) + Q_H(\vartheta)~, \nonumber \\
 N_{AB} (u,\vartheta) \rightarrow N^{D}_{AB}(u,\vartheta) &=& N_{AB} (u - C(\vartheta),\vartheta)~, \nonumber \\
 N^-(\vartheta) \rightarrow N^{-D}(\vartheta) &=& N^-(\vartheta)+Q_H^-(\vartheta)~, \nonumber \\
 N^{-}_{AB} (v,\vartheta) \rightarrow N^{- D}_{AB}(v,\vartheta)&=&N^{-}_{AB} (v-C^-(\vartheta),\vartheta)~. 
 \eea{mp2}
 
We now recognize that this factorization result is absolutely crucial to the preservation of unitarity in scattering processes. Since $C$ and $N$ as defined above are unobservable even in principle, any information leaking into the soft sector would be tantamount to information loss.

We will now turn to defining $C$ and $N$ so that they obey the same algebra but are both observable. We will find that the conservation laws then follow on physical grounds.

\section{$C,N$ as Finite Memories}
\label{sec-cno}

We will now define $C$ and $N$ as observable, finitely-soft gravitational memories. We will show that that satisfy the same algebra as the formal quantitites defined above. The conservation of $C$ and $Q$ can then be understood physically as the trivial propagation of soft degrees of freedom across the spacetime to the antipodal angle.

\subsection{Definitions}

There are actually two reasons why $C$ and $N$ as defined in Sec.~\ref{sec-q} are unobservable. One is that they require access to an infinite time interval. The second is that as a coordinate choice, $C$ would be unobservable even at finite time. It would seem natural, then, to attempt to define $N$ as a finite gravitational memory; and $C$, perhaps, in terms of the positions of physical test masses placed by the observer Bob at large radius, at some finite time.

\begin{figure}[t]
\includegraphics[width=0.8 \textwidth]{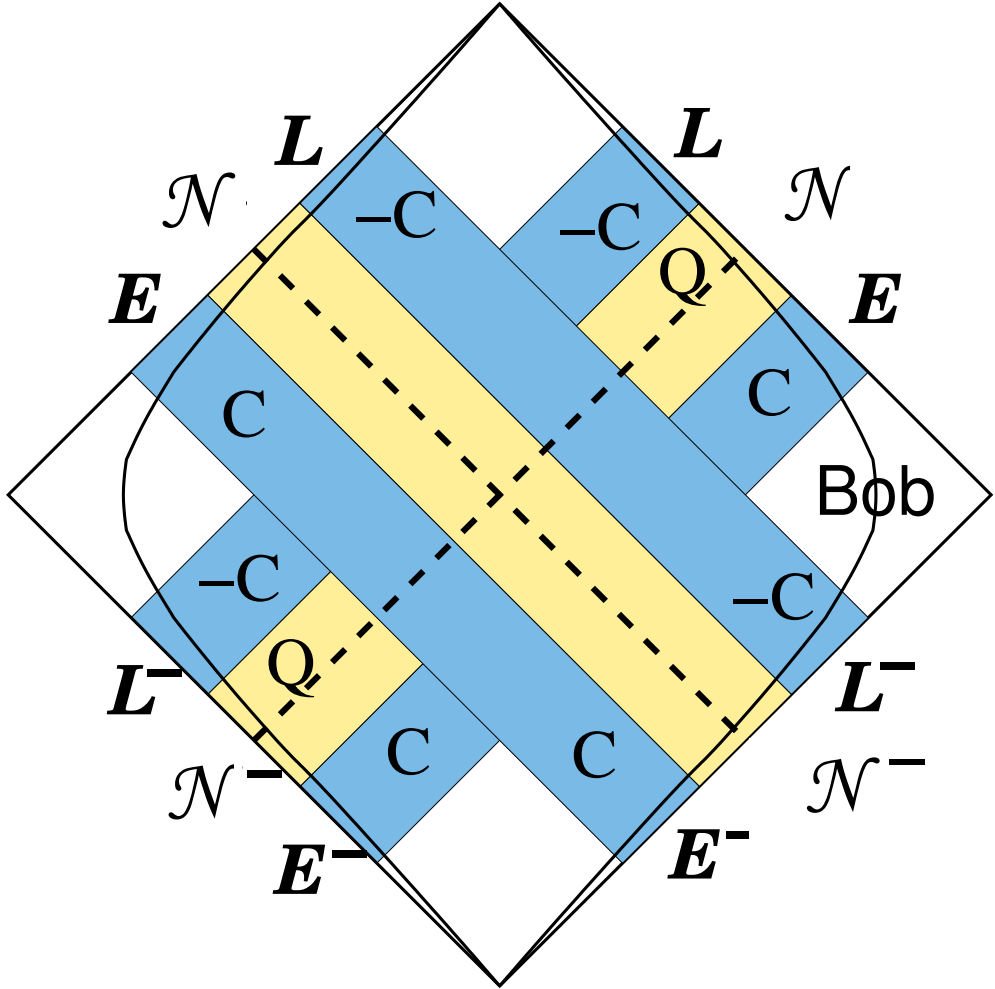}
\caption{``Doubled'' Penrose diagram: every point represents a hemisphere, which makes the antipodal relations visible. The interval ${\cal N}$ (yellow bands) is much greater than the duration of the hard flux (dashed cross). Its (conserved and observable) BMS charge $Q=Q_H+N$ generates observable BMS supertranslations on Bob's equipment. Here this is explicit as $N$ generates (conserved and observable) soft memories $C$ and $-C$ in the intervals $E$ and $L$, which deform Bob's sphere. $Q_H$ generates an angle-dependent time translation that completes this to a BMS transformation.}
\label{fig-sandwich}
\end{figure}

We will indeed define $N$ and $N^-$ as finite memories, measured over a late and an early time interval. The two intervals are taken to be of equal duration, much longer than the greater of the support of the in- and out-going hard radiation. This is shown in Fig.~\ref{fig-sandwich}.

However, we will not define $C$ directly in terms of Bob's test masses. First, it is not clear that the commutator (\ref{eq-cncom}) could then be imposed. Secondly, the antipodal conservation of $C$ would not be natural from this point of view. Absent active intervention, Bob's test masses will follow geodesics at fixed angle, not suddenly switch to an antipodal angle.

Instead we shall take Bob's test masses to have an arbitrary but known position at $t=0$. For definiteness we make the simplest choice,
\begin{equation}
\left. C_{AB}\right|_{t=0}=0~.
\label{eq-ct0}
\end{equation}
(Other choices would introduce a fixed offset in the antipodal identification of $C$ and $C^-$, which is perfectly consistent as well.) 

The question is then how to recover the gauge freedom captured by $C$. We will associate $C$ and $C^-$ with the values of $C_{AB}$ at one end of the interval that defines the memory $N$. We force Bob's test mass positions to reflect these values, by injecting two soft gravitational waves that ``sandwich'' the interval over which $N$ is defined. Their memories interpolate between Eq.~(\ref{eq-ct0}) and the arbitrary $C=C^-$ that we may specify. We will now make this precise.

Consider the advanced or retarded time intervals on ${\cal I}^\pm$ shown in Fig.~\ref{fig-sandwich}. Without loss of generality, we may take the hard radiation to have support well inside the interval $(-1,1)$ in some units. (For example, in the formation and evaporation of a black hole of initial mass $M$, the time unit should be chosen parametrically large in $G^2M^3/\hbar$.) We remove all other radiative degrees of freedom from phase space, except for the soft integrals defined below. Then the hard charge is arbitrarily well approximated by
\begin{eqnarray} 
Q_H[f] & = &  \int d^2 \vartheta\, f\int_{-1}^1 du\, T_{uu}~,\\
Q^-_H[f] & = &  \int d^2 \vartheta\, f\int_{-1}^1 dv\, T_{vv}~.
\end{eqnarray} 

We define the following intervals on ${\cal I}^+$:
\begin{eqnarray} 
{\cal E} & = & \lim_{\epsilon\to 0} (-u_N-\Delta u_C,-u_N+\epsilon)~,\\
{\cal N} & = & (-u_N,u_N)~,\\
{\cal L} & = & \lim_{\epsilon\to 0} (u_N-\epsilon,u_N+\Delta u_C)~.
\end{eqnarray}
Notice that $\epsilon$ is infinitesimal while $\Delta u_C$ is a finite, nonzero interval.
As shown in Fig.~\ref{fig-sandwich}, each of these corresponds to a time interval at positive $t$ on Bob's sphere, via a causal construction. Similarly, we define on ${\cal I}^-$ and at $t<0$ on Bob's sphere the following intervals:
\begin{eqnarray} 
{\cal E}^- & = & \lim_{\epsilon\to 0} (-u_N-\Delta u_C,-u_N+\epsilon)~,\\
{\cal N}^- & = & (-u_N,u_N)~,\\
{\cal L}^- & = & \lim_{\epsilon\to 0} (u_N-\epsilon,u_N+\Delta u_C)~.
\end{eqnarray} 
Note that the constants $u_N, \Delta u_C$ are the same as on ${\cal I}^+$, but the ranges now refer to advanced time, $v$. We require
\begin{equation}
1\ll u_N\ll \Delta u_C \ll r~.
\end{equation}

Thus, the interval ${\cal N}$ is centered on the support of the hard outgoing radiation $N_{AB}$. But ${\cal N}$ is significantly wider, so that it contains most of the long range field of the hard particles. The interval ${\cal E}$ (${\cal L}$) overlaps slightly with ${\cal N}$ but largely precedes (succeeds) it. In this sense the two intervals ``sandwich'' the ${\cal N}$-interval. Analogous statements hold for ${\cal N}^-, {\cal E}^-, {\cal L}^-$ on ${\cal I}^-$.

We now let $\Delta C_{AB}$ denote the gravitational memory accumulated in each interval:
\begin{eqnarray} 
\Delta C_{AB}^{\cal E}(\vartheta) & = & 
\lim_{\epsilon\to 0}\int_{-u_N-\Delta u_C}^{-u_N+\epsilon} du\, N_{AB}(u,\vartheta) 
\label{eq-dcabce}\\
\Delta C_{AB}^{\cal N}(\vartheta) & = & \int_{-u_N}^{u_N} du\, N_{AB}(u,\vartheta)~, \label{eq-dcabn}\\
\Delta C_{AB}^{\cal L}(\vartheta) & = & 
\lim_{\epsilon\to 0}\int_{u_N-\epsilon}^{u_N+\Delta u_C} du\, N_{AB}(u,\vartheta)~.
\label{eq-dcabcl}\\
\Delta C_{AB}^{{\cal E}^-}(\vartheta) & = & 
\lim_{\epsilon\to 0}\int_{-u_N-\Delta u_C}^{-u_N+\epsilon} dv\, N^-_{AB}(v,\vartheta) 
\label{eq-dcabceminus}\\
\Delta C_{AB}^{{\cal N}^-}(\vartheta) & = & \int_{-u_N}^{u_N} dv\, N^-_{AB}(v,\vartheta)~, \label{eq-dcabnminus}\\
\Delta C_{AB}^{{\cal L}^-}(\vartheta) & = & 
\lim_{\epsilon\to 0}\int_{u_N-\epsilon}^{u_N+\Delta u_C} dv\, N^-_{AB}(v,\vartheta)~.
\label{eq-dcabclminus}
\end{eqnarray} 
In analogy with Eqs.~(\ref{eq-n}) and (\ref{eq-vacc}), we now define the soft degrees of freedom $N$ and $C$ as follows: 
\begin{eqnarray} 
N & \equiv & -D^A D^B \Delta C_{AB}^{\cal N}~, \label{eq-nobs}\\
C & \equiv & {1\over D^2 (D^2+2)}D^AD^B \Delta C_{AB}^{\cal E}
\label{eq-vacdc}
\end{eqnarray} 
Analogously we define at early times: 
\begin{eqnarray} 
N^- & \equiv & -D^A D^B \Delta C_{AB}^{\cal N^-}~, \label{eq-nobs}\\
C^-&=& {1\over D^2 (D^2+2)}D^AD^B \Delta C_{AB}^{{\cal L}^-}
\label{eq-vacdcm}
\end{eqnarray} 
The operator $D^2(D^2+2)$ is equal to $(l-1)l(l+1)(l+2)$ on the $l$-th spherical harmonics. Since it vanishes for $l=0,1$, the proper definition of $C$ and $C^-$ requires expanding $D^AD^B \Delta C_{AB}^{\cal E}$ and $D^AD^B \Delta C_{AB}^{{\cal L}^-}$ in spherical harmonics, and projecting out the $l=0,1$ coefficients. 


We have not yet made use of the two remaining memories introduced above: the early memory on ${\cal I}^-$, and the late memory on ${\cal I}^+$. They will be fixed by the symplectic structure we wish to achieve, in the next subsection.

\subsection{Derivation of the $\{N,C\}$ Commutator}

We can now derive, rather than merely impose, the commutation relations satisfied by $C$ and $N$. To do this, we integrate the Dirac bracket of the Bondi news~\cite{Ash81,Wal11}: 
\begin{equation}
\{N_{AB}(u,\vartheta),N^{CD}\!(u'\!, \vartheta')\} = 16\pi G\, \delta_{AB}^{CD}
\partial_u \delta (u-u')\delta^2(\vartheta -\vartheta')
\label{eq-nabnab}
\end{equation}
where $ \delta_{AB}^{CD}\equiv \delta_A^C\delta_B^D+\delta_A^D\delta_B^C-h_{AB} h^{CD}$. 

Taking $-1<u'<1$ and integrating $u$ over either ${\cal E}$ or ${\cal N}$ shows that $C$ and $N$ both commute with the hard degrees of freedom $N_{AB}(u',\vartheta)$, as required. In particular, the hard charge $Q_H$ commutes with the soft degrees of freedom as required:
\begin{equation}
\{N,Q_H\}=0=\{C,Q_H\}~.
\end{equation}

Next, let us integrate Eq.~(\ref{eq-nabnab}) twice: first over $u$ in the ${\cal N}$ interval, then over $u'$ in the ${\cal E}$ interval. This yields
\begin{equation}
\{\Delta C_{AB}^{\cal N}(\vartheta),\Delta C^{CD}_{\cal E}(\vartheta')\} = 
-16\pi G\,  \delta_{AB}^{CD}\, \delta^2(\vartheta -\vartheta')~.
\label{eq-cabcab}
\end{equation}
Recalling that on a scalar function $\Phi$ covariant derivatives on the unit sphere obey $(D_A D_B D_A  - D_B D^2) \Phi= D_B\Phi$, it is now a matter of simple algebra to check that our variables $C,N$ obey the standard canonical commutator that is usually imposed by hand:
\begin{equation}
\{N(\vartheta),C(\vartheta')\} = 16\pi G\, \delta^2(\vartheta-\vartheta')~.
\label{eq-cncom2}
\end{equation} 
Similarly one finds on ${\cal I}^-$ that 
\begin{equation}
\{\Delta C_{AB}^{{\cal N}^-}(\vartheta),\Delta C^{CD}_{{\cal L}^-}(\vartheta')\} = 
-16\pi G\,  \delta_{AB}^{CD}\, \delta^2(\vartheta -\vartheta')~,
\label{eq-cabcabminus}
\end{equation}
whence
\begin{equation}
\{N^-(\vartheta),C^-(\vartheta')\} = 16\pi G\, \delta^2(\vartheta-\vartheta')~.
\label{eq-cncom2minus}
\end{equation} 

Definition (\ref{eq-vacdc})  selects one particular component of the memory 
$\Delta C^{CD}_{\cal E}$, so, in a general setting, Bob would need to know the other
component to determine the displacement of his detectors at $u=-1$. In our setup, only soft radiation arrives in the 
${\cal E}$ interval. This is essentially the non-radiative condition of \cite{FlaNic15}, which implies that the memory assumes the
special form 
\beq
\Delta C_{AB}^{\cal E}(\vartheta) = (-2D_A D_B + h_{AB}D_C D^C)\, C(\vartheta)~,
\eeq{non-radiat}
with only one independent component. An analogous property holds on ${\cal I}^-$, so in our setting the memories $C,C^-$ are all that Bob needs to set up properly his experiment.

\subsection{Symplectic Structure}
\label{sec-symp}

One more step is needed to obtain a consistent symplectic structure. So far, we have not associated the early memory on ${\cal I}^-$ with any conjugate variable. Similarly on ${\cal I}^+$, the late memory on ${\cal I}^+$ lacks a conjugate. Roughly speaking, we have three soft variables at every angle. But we need phase space to be even-dimensional. 

Moreover, we would like $N$ to commute with everything but $C$. But by the same derivation that led to Eq.~(\ref{eq-cabcab}), one finds that  
\begin{eqnarray} 
\{\Delta C_{AB}^{\cal N}(\vartheta),\Delta C^{CD}_{\cal L}(\vartheta')\} & = &
16\pi G\,  \delta_{AB}^{CD}\, \delta^2(\vartheta -\vartheta')~,
\label{eq-cabcab2}\\
\{\Delta C_{AB}^{{\cal N}^-}(\vartheta),\Delta C^{CD}_{{\cal E}^-}(\vartheta')\} & = & 
16\pi G\,  \delta_{AB}^{CD}\, \delta^2(\vartheta -\vartheta')~,
\label{eq-cabcab2minus}
\end{eqnarray} 
so $N$ has a nonzero commutator with the unpartnered memory.

We can resolve both of those shortcomings by identifying the early memory with {\em minus} the late memory on ${\cal I}^+$, up to an arbitrary constant that we set to zero for convenience.\footnote{This constant could be chosen nonzero, just as we could choose a nonzero constant in Eq.~(\ref{eq-ct0}). There is nothing fundamentally significant about any particular choice of gauge, and a well-posed scattering problem remains well-posed whether or not an offset is introduced between $C$ and $C^-$.} The minus sign is necessary to accommodate the relative sign in Eqs.~(\ref{eq-cabcab}) and (\ref{eq-cabcab2}). By the same argument, we also relate the early and late memory on ${\cal I}^-$:
\begin{eqnarray} 
\Delta C_{AB}^{\cal E}(\vartheta) & = & 
-\Delta C_{AB}^{\cal L}(\vartheta)~, \label{eq-elp} \\
\Delta C_{AB}^{{\cal E}^-}(\vartheta) & = & 
-\Delta C_{AB}^{{\cal L}^-}(\vartheta)~. \label{eq-elm}
\end{eqnarray}

Note that we are not saying that these memories must be so related in Nature. Of course, they can be dialed independently. Rather, our goal is to reproduce the bracket algebra and conservation laws prevalent in the recent literature, but using degrees of freedom that are all observable in finite time. We are free to identify or eliminate any degrees of freedom as long as this is consistent with the dynamics and as long as the algebra remains closed. We have assured the latter by including a minus sign in the above identification. 

Next we will turn to the dynamics. We will show that the desired conservation laws emerge automatically.

\subsection{Conservation of $Q$ and $C$}

We now make the physical observation that soft gravitons propagate trivially through the spacetime, without scattering off of one another or interacting with the hard radiation. This is the reason why, for example, the cosmic microwave background does not affect experiments in particle accelerators. This simple fact will lead us to two conservation laws that have been asserted for the unobservable versions of $C$ and $Q$: namely, that $C=C^-$, and that $Q=Q^-$. 

The first of these is obvious: the failure of the early memory to scatter implies that it is the same on ${\cal I}^+$ and on ${\cal I}^-$. Similarly, the late memory is also conserved. We can readily see this from Fig.~\ref{fig-sandwich}. Recall the antipodal identification of angles between ${\cal I}^\pm$; this is now a convenient choice, because soft gravitons that enter at one angle will exit through the opposite angle. We thus have
\begin{eqnarray}
\Delta C_{AB}^{{\cal E}^-}(\vartheta) & = & 
\Delta C_{AB}^{{\cal E}}(\vartheta)~, \label{ccons1} \\
\Delta C_{AB}^{{\cal L}^-}(\vartheta) & = & 
\Delta C_{AB}^{\cal L}(\vartheta) ~. \label{ccons2}
\end{eqnarray} 
From Eqs.~(\ref{eq-vacdc}), (\ref{eq-vacdcm}), (\ref{eq-elp}), and (\ref{eq-elm}), we can conclude that
\begin{equation}
C=C^-~.
\end{equation}

To see that $Q$ is conserved, we recall that $Q$ is simply the Bondi mass aspect, $m_B$~\cite{Ash81,Str13}, at the retarded time $-u_N$ or Bob's clock time $t^+=r-u_N$; and $Q^-$ is the Bondi mass aspect at the advanced time $u_N$ or $t^-=-r+u_N$. In the case of Maxwell theory, $Q$ and $Q^-$ are given by the components $F_{rt}$ of the Maxwell field at these two times. Here we complete the argument for the Maxwell case, where $C$ and $C^-$ can be similarly defined as observable memories in the ${\cal E}$ and ${\cal L}$ intervals (see e.g.~\cite{Str17,BouPor17} for formulas for the electromagnetic memory). The gravitational case is analogous.

We first consider the case where $C=0$, so that there is no radiation, not even soft radiation, outside the intervals ${\cal N}$ and ${\cal N}^-$ on Bob's sphere.

We take the radius $r$ of Bob's sphere so large that the entire hard scattering process can be treated as occuring at the event $r=t=0$. Bob's whole sphere is spacelike to this point at the two times $t^\pm$, and at all intermediate times. Bob's world tube thus surrounds a collection of distant, freely propagating charged particles that all reach $r=0$ at the time $t=0$. By causality, the asymptotic field strength $F_{rt}$ must be given by a linear superposition of Li\'enard-Wiechert solutions for such charges (see, e.g., Ref.~\cite{Str17}).   

The solution is antipodally symmetric under $t\to -t$ at fixed $r$. In our convention where fixed $\vartheta$ is already understood to flip to the antipodal position on the sphere as $t=0$ is crossed, this becomes the statement that
\begin{equation}
F_{rt}(t,\vartheta)=F_{rt}(-t,\vartheta)~,~~ -r+u_N\leq t\leq r-u_N~.
\end{equation}
As discussed above, the choice $t=r-u_N$ yields $Q=Q^-$. 

The Li\'enard-Wiechert solution is in general valid for massive particles only.  For massless particles moving on a general light-like trajectory the solution is subtle. For the same reason as in the massive case, we need it only for particles moving on a straight line. In this case the solution is simple and is also antipodally symmetric under $(t,r) \rightarrow (-t,r)$. See Eq.~(1.1) of~\cite{lechner} and references therein.

We can now include the effects of a nonzero value of $C$. In our implementation, this means that nonzero soft memories are present in the early and late intervals; see Eqs.~(\ref{eq-vacdc}) and (\ref{eq-vacdcm}). We now replace $u_N\to u_N+\Delta u_C$ in applying the above argument. This ensures that the Bondi mass aspect, or $F_{rt}$ in the Maxwell case, at the end of the ${\cal L}^-$ interval agrees antipodally with the same quantity at the beginning of the ${\cal E}$ interval. In the absence of hard flux, the mass aspect or $F_{rt}$ change precisely by the soft memory accumulated in an interval (see, e.g., Ref.~\cite{Str17}). By Eqs.~(\ref{eq-elp}), and (\ref{eq-elm}), the memory in ${\cal L}^-$ is equal and opposite to that of ${\cal E}$. Thus $F_{rt}$ or $m_B$ still agrees antipodally at $t^\pm$, where they define the charge $Q$ and $Q^-$ respectively. Hence 
\begin{equation}
Q=Q^-~.
\end{equation}

\paragraph*{Acknowledgments} R.B.\ was supported in part by the Berkeley Center for Theoretical Physics, by the National Science Foundation (award numbers PHY-1521446, PHY-1316783), by FQXi, and by the US Department of Energy under contract DE-AC02-05CH11231.  M.P.\ was supported in part by NSF grant PHY-1620039. M.P.\ would like to thank the LPTENS 
for its kind hospitality during the completion of this work.





\bibliographystyle{utcaps}
\bibliography{all}
\end{document}